\documentclass[aps,prc,twocolumn,showpacs,floatfix,nofootinbib,preprintnumbers,superscriptaddress,amsmath,amssymb]{revtex4-1}

\usepackage{graphicx}
\usepackage{epsfig}
\usepackage{bm}
\usepackage{color}
\usepackage{float}
\usepackage{dcolumn}
\usepackage{multirow} 
\usepackage{hyperref}

\graphicspath{{.}{../}}

\newcommand{\beq}{\begin{equation}}
\newcommand{\eeq}{\end{equation}}
\newcommand{\beqa}{\begin{eqnarray}}
\newcommand{\eeqa}{\end{eqnarray}}

\begin{document}

\title{Ab-initio coupled-cluster effective interactions for the shell model:
Application to neutron-rich oxygen and carbon isotopes}

\author{G.~R.~Jansen} \affiliation{Physics Division, Oak Ridge National
Laboratory, Oak Ridge, TN 37831, USA}
\affiliation{Department of Physics and Astronomy, University of Tennessee,
Knoxville, TN 37996, USA}

\author{J.~Engel} \affiliation{Deptartment of Physics and Astronomy, University
of North Carolina, Chapel Hill, NC, 27516-3255, USA}

\author{G.~Hagen} \affiliation{Physics Division, Oak Ridge National Laboratory,
Oak Ridge, TN 37831, USA} 
  \affiliation{Department of Physics and Astronomy, University of Tennessee,
  Knoxville, TN 37996, USA} 

\author{P.~Navratil} \affiliation{TRIUMF, 4004 Wesbrook Mall, Vancouver,
British Columbia, V6T 2A3 Canada}

\author{A.~Signoracci} \affiliation{Physics Division, Oak Ridge National
Laboratory, Oak Ridge, TN 37831, USA}
\affiliation{Department of Physics and Astronomy, University of Tennessee,
Knoxville, TN 37996, USA}

\begin{abstract}
  We derive and compute effective valence-space shell-model
  interactions from ab-initio coupled-cluster theory and apply them to
  open-shell and neutron-rich oxygen and carbon isotopes. Our
  shell-model interactions are based on nucleon-nucleon and
  three-nucleon forces from chiral effective-field theory. We compute
  the energies of ground and low-lying states, and find good agreement
  with experiment. In particular our calculations are consistent with
  the $N=14, 16$ shell closures in $^{22,24}$O, while for $^{20}$C the
  corresponding $N=14$ closure is weaker. We find good agreement
  between our coupled-cluster effective-interaction results with those
  obtained from standard single-reference coupled-cluster calculations
  for up to eight valence neutrons.
\end{abstract}

\pacs{21.30.Fe, 21.60.Cs, 21.60.De, 21.10.-k}

\maketitle {\it Introduction.} -- The nuclear shell model is the
foundation on which our understanding of nuclei is built.
One of the most important problems in nuclear structure
today is to understand how shell structure changes with
neutron-to-proton ratio throughout the nuclear chart. Shell structure
influences the locations of the neutron and proton drip lines and the
stability of matter. Examples of changes in shell structure are the
appearance of new magic numbers $N=14$ and $N=16$ in the neutron-rich
oxygen isotopes \cite{stanoiu2004,kanungo2009}, and the emergence of
an $N=34$ sub-shell closure in $^{54}$Ca
\cite{holt2012,hagen2012b,steppenbeck2013,steppenbeck2013b}.

Phenomenological shell-model Hamiltonians such as the $sd$ Hamiltonian of Brown
and Wildenthal \cite{wildenthal1984,brown1988} (abbreviated USD) and
the $p$-$sd$ Hamiltonian of Warturburton and Brown \cite{warburton1992} (abbreviated
WBP), have successfully described properties of nuclei with proton number $Z$
and neutron number $N$ less than about 20.  To understand the origin of shell
structure, however, researchers are now trying to derive the shell model from
realistic nucleon-nucleon (NN) and three-nucleon forces (3NFs), without
further phenomenology \cite{hjorthjensen1995,otsuka2010,holt2012}.  Within the
last few years, for example, ~\textcite{otsuka2010} showed that 3NFs play a
pivotal role in placing the drip line (correctly) in the oxygen isotopes at
$^{24}$O, and ~\textcite{holt2012} showed that the inclusion of 3NFs are
necessary to explain the high $2^+$ state in $^{48}$Ca .  

Until recently, all work to compute effective shell-model interactions was
perturbative.  Lately, however, nonperturbative calculations have become
possible.  Some have been based on the ab-initio no-core shell model
\cite{navratil2009,barrett2013}, via a valence-cluster expansion
\cite{navratil1997,lisetskiy2008,shukla2011}, and others on the in-medium
similarity renormalization group \cite{tsukiyama2012}. In this Letter we use a
third approach, the ab-initio coupled-cluster method
\cite{coester1958,coester1960,cizek1966,cizek1969,kuemmel1978,hagen2013c}, to construct
effective shell-model interactions for use in open-shell and neutron-rich
nuclei. Starting from NN interactions and 3NFs generated by chiral
effective-field-theory, we compute the ground- and excited-state energies of
neutron-rich carbon and oxygen isotopes with up to eight neutrons in the valence
space.  Intense theoretical and experimental interest surround the structure of
both these isotope chains, and particularly the neutron-rich carbon isotopes.
Separation energies, spin assignments for low-lying states, the energies of
$2^+$ states, and transition rates in these isotopes all depend on the
locations of shell gaps.
\cite{kondo2009,strongman2009,ozawa2011,kobayashi2012,petri2012,voss2012,petri2011,forssen2013}.
At present there is no evidence for a shell closure at the $N=14$ nucleus
$^{20}$C \cite{stanoiu2008}, despite the $N=14$ shell closure at $^{22}$O.
Furthermore, Efimov physics may be at play in $^{22}$C
\cite{tanaka2010,ershov2012}. This Letter takes the first steps towards an
ab-initio shell-model description of the neutron-rich carbon isotopes, and
addresses the role of 3NFs in these isotopes.

{\it Hamiltonian and model space.} -- Our coupled-cluster calculations start
from the intrinsic $A$-nucleon Hamiltonian,
\begin{equation}
  \label{intham}
  \hat{H} = \sum_{i<j}\left({({\bf p}_i-{\bf p}_j)^2\over 2mA} + \hat{V}
    _{NN}^{(i,j)}\right) + \sum_{ i<j<k}\hat{V}_{\rm 3N}^{(i,j,k)}.
\end{equation}
Here the intrinsic kinetic energy (the first term) depends on the mass number
$A \equiv Z+N$.  The potential $\hat{V}_{NN}$ denotes the chiral NN interaction
at next-to-next-to-next-to leading order~\cite{entem2003,machleidt2011} (with
cutoff $\Lambda=500$ MeV), and $\hat{V}_{\rm 3N}$ is the 3NF that enters at
next-to-next-to leading order with a local regulator \cite{navratil2007} (with
cutoff $\Lambda_{\rm 3N} = 400$ MeV). The low-energy constants of the 3NF are
given by $c_E = 0.098$ and $c_D = -0.2$. These were initially determined from a
fit to the triton half-life and binding energy with a cutoff $\Lambda_{\rm 3N}
= 500$ MeV \cite{gazit2009}, and then, with $\Lambda_{\rm 3N} = 400$ MeV, $c_E$
was re-adjusted to reproduce the $^4$He binding energy while $c_D$ was kept
fixed \cite{roth2012}.  To achieve faster convergence as model space size is
increased, we use the similarity renormalization group (SRG) to evolve
$\hat{V}_{\rm NN}$ and $\hat{V}_{\rm 3N}$ to the lower momentum scale
$\lambda_{\rm SRG} = 2.0\ \mathrm{fm}^{-1}$ \cite{jurgenson2009}.  For the
coupled-cluster calculations themselves, we use a Hartree-Fock basis, with each
orbital expanded in another basis containing up to $N_{\rm max} + 1 = 13 $ major
shells from a harmonic-oscillator potential with frequency $\hbar\omega =
20$MeV. We limit the number of 3NF matrix elements through the additional cut
$N_1 + N_2 + N_3 \leq 14$. The resulting model-space is sufficient to obtain
well converged results for energies of the states reported in this Letter. We
use the normal-ordered two-body approximation for the 3NF
\cite{hagen2007a,roth2012}, which has been shown to work well in light-
and medium mass nuclei \cite{roth2012}.  By using a Hartree-Fock basis we
minimize the oscillator-frequency dependence arising from the neglect of
residual 3NFs.

{\it Formalism.} -- To derive an effective shell-model Hamiltonian in a valence
space from ab-initio coupled-cluster theory, we use the valence-cluster
expansion first applied in the no-core shell model 
\cite{navratil1996,lisetskiy2008}. We expand the Hamiltonian in
Eq.~\eqref{intham} in a form suitable for the shell model:
\begin{equation}
  \label{ham}
  H^A_{\rm CCEI} = H_0^{A,A_c} + H_1^{A,A_c+1} + H_2^{A,A_c+2} + \ldots,
\end{equation}
where CCEI stands for coupled-cluster effective interaction, $A$ is
the mass of the nucleus we wish to treat, and $A_c$ is the mass of the
nucleus with a closed core below the valence space. In
Eq.~\eqref{ham}, $H_0^{A,A_c} $ is the Hamiltonian for the core,
$H_1^{A,A_c+1}$ is the valence one-body Hamiltonian, and
$H_2^{A,A_c+2}$ is the additional two-body piece. In this work we
limit ourselves to one- and two-body terms in the valence-space
shell-model Hamiltonian. To solve for the ground-state of the core
nucleus $A_c$ we use the coupled-cluster method in the
singles-and-doubles approximation (CCSD) with the $\Lambda$-triples
correction treated perturbatively ($\Lambda$-CCSD(T))
\cite{kucharski1998,taube2008}. To obtain the ground and excited
states for the $A_c+1$ and $A_c+2$ nuclei we use the one- and
two-particle-attached equation-of-motion (EOM) coupled-cluster methods
\cite{gour2006, jansen2011, jansen2012}.  For the particle-attached
EOM we truncate at one-particle and two-particle-one-hole excitations,
and for the two-particle-attached EOM we truncate at two-particle and
three-particle-one-hole excitations. In coupled-cluster theory the
basic ingredient is the similarity-transformed Hamiltonian
$\overline{H} = e^{-T}He^T$, which is inherently non-Hermitian
\cite{bartlett2007}, thus we need to solve for the left- and the right
eigenstates to obtain a complete \textit{bi}-orthogonal set of
states. From the left and right eigen-states we can write
Eq.~\eqref{ham} in a spectral representation.

The valence space representation of Eq.~\eqref{ham} consists of a core
energy term (which we compute from $H_0^{A,A_c}$ using
$\Lambda$-CCSD(T)), a one-body term (built from the particle-attached
eigenvalues of $H_1^{A,A_c+1}$), and a two-body term. The two-body
term is computed using the Okubo-Lee-Suzuki similarity transformation
\cite{okubo1954,suzuki1980,suzuki1982,kvaal2008} by projecting the
two-particle attached EOM eigen states onto two-body valence-space
states. The Okubo-Lee-Suzuki
projection of $H_2^{A,A_c+2}$ onto the model-space is
\cite{navratil1996,kvaal2008},
\begin{equation}
\label{eq:simtrans}
\langle \alpha _P\vert \overline{H_{\rm eff}^{A}} \vert \alpha _{P'} \rangle = 
\sum_{k=1}^d \langle \alpha _P\vert  R_k^{A,A_c+2}\rangle e_k
\overline{ \langle \alpha_{P'}\vert  R_k^{A,A_c+2}\rangle}. 
\end{equation}
Here the $\vert R_k^{A,A_c+2}\rangle $ are the two-particle attached
EOM eigenstates with eigenvalue $e_k$ for $A_c+2$ (with mass $A$ of
the target nucleus in the kinetic energy), $ \vert \alpha _ P\rangle $
are the model-space states, the sum is over the $k$ two-particle
attached eigenstates that have the largest overlap with the model
space. The $ \langle \alpha _P\vert R_k^{A,A_c+2}\rangle $ are the
matrix elements of the unitary operator ${\bf U}$. $\overline{ \langle
  \alpha_{P'}\vert R_k^{A,A_c+2}\rangle} $denote the matrix elements
of the inverse of ${\bf U}$. To obtain the effective two-body
shell-model interaction, we subtract the one-body part from
Eq.~\eqref{eq:simtrans} to avoid double counting.

Note, that we could also construct an effective Hamiltonian using the
corresponding left eigenvectors. This introduces an ambiguity in the
construction of $\overline{H_{\rm eff}^{A}}$. We have verified,
however, that in this work the matrix elements of the effective
operator are almost identical for either choice. The effective
Hamiltonian in Eq.~\eqref{eq:simtrans} is not Hermitian. Current
shell-model codes require Hermitian matrices. To obtain a Hermitian
representation of the effective shell-model Hamiltonian we construct
the metric operator $S^\dagger S$ where $S$ is a matrix that
diagonalizes $ {H}^A_{\rm CCEI} $; the Hermitian shell-model
Hamiltonian is then $\left[S^\dagger S\right] {H}^A_{\rm CCEI}
\left[S^\dagger S\right]^{-1/2}$ \cite{scholtz1992,navratil1996}.

{\it Results.} -- Here we report our CCEI results for ground and low-lying
states in oxygen and carbon isotopes. We choose $^{16}$O in oxygen and $^{14}$C
in carbon as the closed-shell cores.  We then project the one- and
two-particle-attached coupled-cluster wave functions onto the one- and
two-particle model space states in the valence space --- the
$d_{5/2},s_{1/2},d_{3/2}$ shell --- and proceed to use the resulting
shell-model Hamiltonians in heavier isotopes.

\begin{figure}[b!]
  \centerline{\includegraphics[width=1.0\columnwidth]{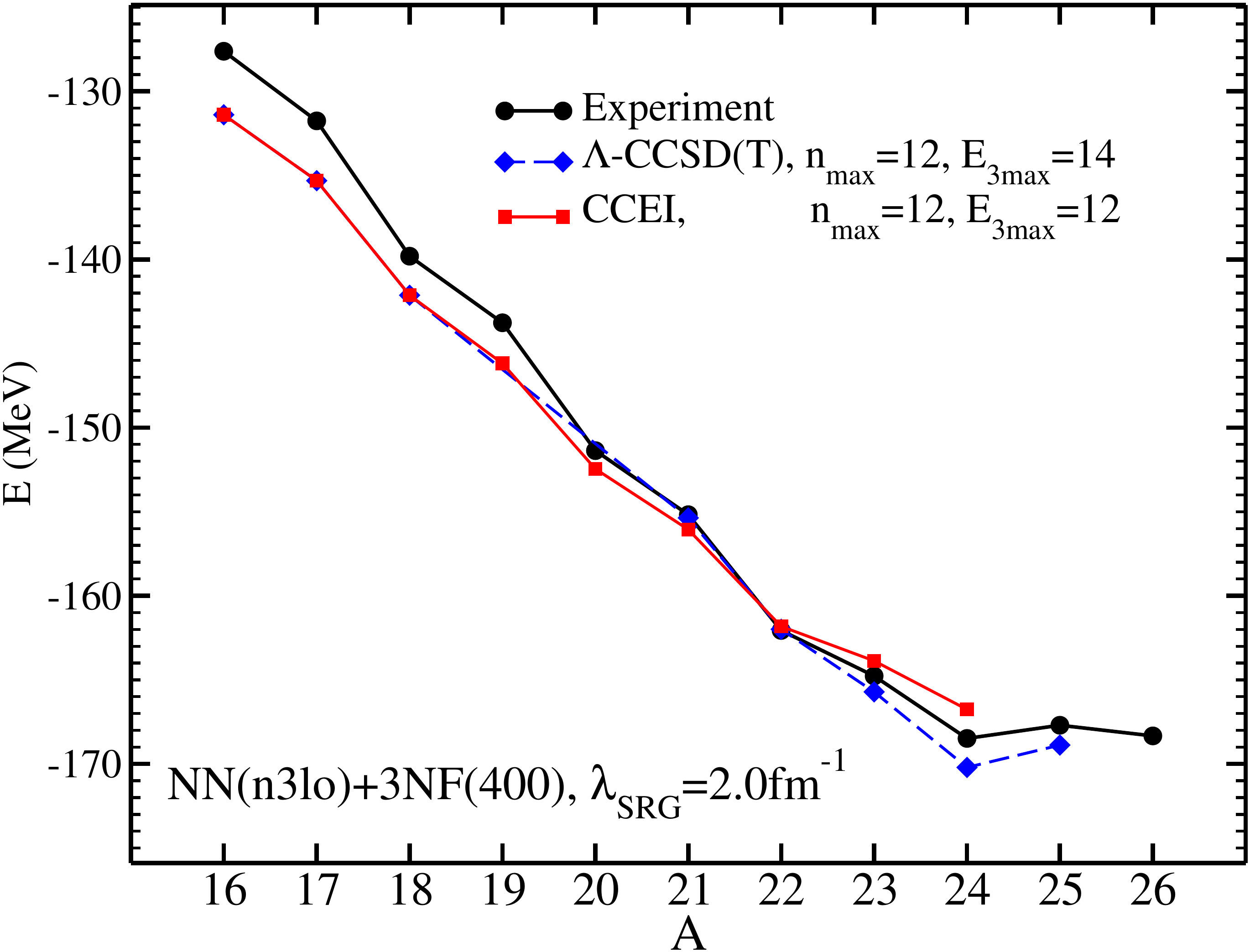}}
  \vspace*{0.0cm}
  \caption{(Color online) Ground-state energies of oxygen isotopes. Black
  circles show the experimental values, blue diamonds the $\Lambda$-CCSD(T)
  results, and the red squares the CCEI results for the ground-state energies.}
  \label{be_oxygen}
\end{figure}

\begin{figure*}[tbh]
\includegraphics[width=0.8\textwidth]{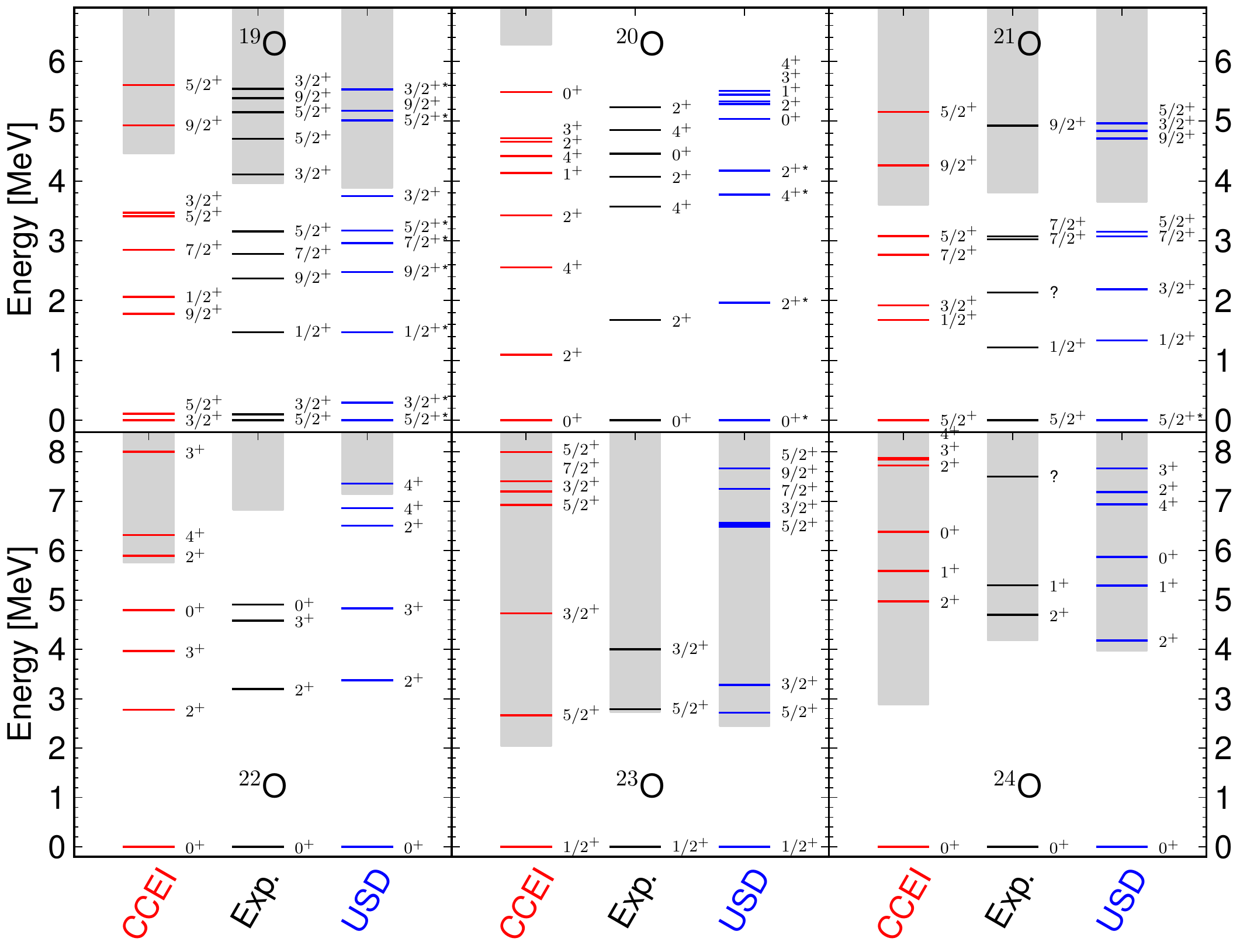}
\caption{(Color online) Excitation spectra of neutron-rich oxygen isotopes.
The left columns (red lines) contain the CCEI results, the middle columns
(black lines) the known experimental data, and the right columns (blue lines)
the spectra obtained with the USD shell-model Hamiltonian
\cite{wildenthal1984,brown1988}.  A star next to the excitation levels in the
right columns indicates that the level was included in the fit of the USD
Hamiltonian.  The gray bands indicate states above the neutron decay threshold. }
  \label{ex_oxygen}
\end{figure*}

\begin{figure*}[tbh]
\includegraphics[width=0.8\textwidth]{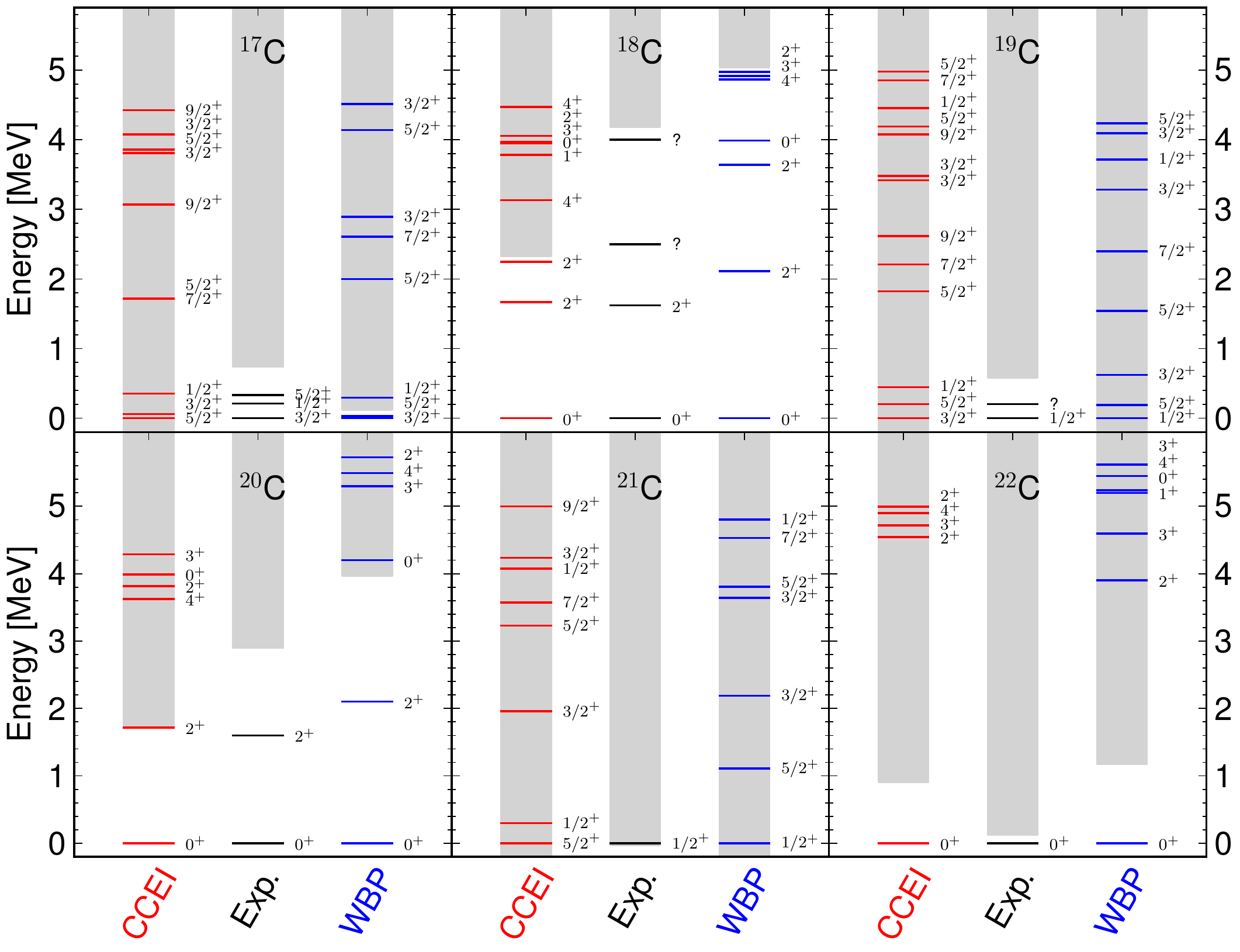}
\caption{(Color online) Same caption as in Fig.~\ref{ex_oxygen}
  except for carbon isotopes. The right column (blue lines) show the spectra
  obtained with the WBP shell-model Hamiltonian \cite{warburton1992}.}
  \label{ex_carbon}
\end{figure*}

We would like to gauge the accuracy of our CCEI approach by comparing
with full $\Lambda$-CCSD(T) calculations, the results of which we
refer to as reference values. Figure~\ref{be_oxygen} shows the
ground-state energies of all oxygen isotopes $^{16-24}$O computed with
the CCEI (red squares), experimental ground-state energies (black
circles), and the $\Lambda$-CCSD(T) ground-state energies in
$^{21,22,23,24}$O. Our $\Lambda$-CCSD(T) calculations use the model
space mentioned earlier, while the calculations that determine our
CCEI use $N_{\rm max} =12 $ and $N_1+N_2+N_3 = 12$. We believe that
our CCEI results are converged to within $\sim 100$~keV.  Both our
$\Lambda$-CCSD(T) and CCEI results are in good agreement with
experimental binding energies. Our CCEI and $\Lambda$-CCSD(T)
calculations also agree well with a variety of recent calculations in
the oxygen isotopes that start with the same Hamiltonian
\cite{hergert2013,cipollone2013}.

If we look more closely, we see that the reference $\Lambda$-CCSD(T) results in
$^{21,22}$O are in excellent agreement with our CCEI results. In $^{23,24}$O
the CCEI results start to deviate from the $\Lambda$-CCSD(T) reference values.
In $^{24}$O the CCEI ground-state is less bound by about $3.5$~MeV than
obtained with $\Lambda$-CCSD(T). The difference indicates that effective
three-body interactions induced by the Okubo-Lee-Suzuki transformation (which
we neglect) start to play a role in the CCEI approach when the number of
valence nucleons gets too large. The problem can be remedied by including these
interactions or by increasing the valence space size. 

Next, we compare low-lying CCEI excited-state energies in $^{22}$O
with an EOM coupled-cluster calculation that includes singles and
doubles excitations \cite{stanton1993}. EOM-CCSD can accurately
describe low-lying states that are dominated by one-particle-one-hole
excitations \cite{bartlett2007}, and we therefore choose those states
for comparison.  In $^{22}$O we obtain low-lying $2^+$ and $3^+$
states with $2.5$~MeV and $3.5$~MeV of excitation energy. The CCEI
result for the same states is $2.7$~MeV and $4.0$~MeV, though the CCEI
result for the $3^+$ state in $^{22}$O is not yet converged; it moves
down by $\sim 150$~keV when we increase the model space size from
$N=10$ to $N=12$ oscillator shells.  The $2^+$ state changes only by
$\sim 5$~keV indicating that it, by contrast, is well
converged. Standard EOM-CCSD works well for states that are dominated
by one-particle-one-hole excitations. In our CCEI calculations,
correlations between all particles in the valence space are treated
exactly. Therefore, we expect to see some differences in the computed
spectra. For example, in CCEI we are able to compute the second $0^+$
state in $^{22}$O, which is dominated by two-particle-two-hole
excitations from the ground-state.

We turn now to carbon.  The $\Lambda$-CCSD(T) ground-state energies of
$^{14,15,16}$C are $-104.0$~MeV, $-104.2$~MeV, and $-106.6$~MeV,
respectively.  In $^{14}$C the result agrees well with the
experimental ground-state energy of $-105.3$~MeV, but for $^{15,16}$C
our particle-attached and two-particle-attached EOM results are
$2.3$~MeV and $4.2$~MeV under-bound with respect to experimental data.
The under-binding persists throughout the chain of carbon isotopes in
our CCEI calculations.

Figures ~\ref{ex_oxygen} and ~\ref{ex_carbon} summarize our CCEI results for
the excited states in the neutron-rich oxygen and carbon isotopes. Figure
~\ref{ex_oxygen} shows the excitation spectra for $^{19-24}$O.  Our results are
overall in very good agreement with the experimental excited-state energies.
And without any adjustment of parameters we obtain spectra that are
qualitatively similar to that produced by the phenomenological USD
Hamiltonian.  Our $2^+$ states in the even oxygen isotopes are on
target and consistent with $N=14$ sub-shell closure in $^{22}$O and the $N=16$
shell closure in $^{24}$O.

Figure ~\ref{ex_carbon} shows excitation spectra in $^{17-22}$C. Here
the right column contains results produced by the phenomenological
$p$-$sd$ WBP shell-model Hamiltonian \cite{warburton1992} with
$2\hbar\omega$ excitations. As in the oxygen isotopes we agree very
well overall with experiment, and our $2^+$ states are remarkably
close to the data. Our $2^+$ energy in $^{20}$C is at $1.72$~MeV while
the corresponding $2^+$ state in $^{22}$O lies at $2.78$~MeV; thus our
results are consistent with a weaker $N=14$ sub-shell closure in the
carbon isotopes.

In the odd isotopes $^{17,19,21}$C we get the $1/2^+$ state in the wrong
position.  Our calculations, however, rely on an underlying harmonic-oscillator
basis and therefore do not account for the particle continuum.  The $1/2^+$
state is dominated by $s$-waves and is located close to the particle emission
threshold, where continuum effects are obviously important
\cite{hagen2012a,hagen2012b}.  The $3/2^+$ and the $5/2^+$ states are dominated
by $d$-waves, which couple somewhat less to the continuum because of the $l=2$
centrifugal barrier \cite{hagen2012a}.  Overall, we expect continuum effects to
be quite significant for the $^{17,19,21}$C isotopes.  Preliminary calculations
within the no-core shell model with continuum \cite{baroni2013b} for $^{17}$C,
with the same chiral NN+3N interaction used here, show that the $1/2^+$ state
(unbound in our calculation) gains about 2 MeV in energy and becomes bound. At
the same time, the $3/2^+$ and $5/2^+$ states are lowered in energy by more
than 1 MeV.  We anticipate similar or even stronger continuum effects $^{19}$C
and $^{21}$C; these would most likely make the $1/2^+$ states the ground
states, as they are in reality. 

{\it Summary.} -- We have used coupled cluster theory to derive
shell-model Hamiltonians that depend on no parameters other than those
in the initial NN and $3N$ chiral interactions.  We have reproduced
ground and excitation energies with good accuracy in carbon and oxygen
isotopes. The results demonstrate both the predictive power of
Hamiltonians from chiral EFT and the accuracy of the coupled cluster
framework.  Finally, our shell model calculations can and will be
systematically improved (e.g.\ by including induced three-body
interactions), extended to include effective operators other than the
Hamiltonian, and applied to heavier nuclei, where accurate
phenomenological Hamiltonians are harder to obtain.

\begin{acknowledgments}
  We thank Scott Bogner, Morten Hjorth-Jensen and Thomas Papenbrock
  for useful discussions. This work was supported by the Office of
  Nuclear Physics, U.S. Department of Energy (Oak Ridge National
  Laboratory), under DE-FG02-96ER40963 (University of Tennessee),
  DE-FG02-97ER41019 (University of North Carolina), DE-SC0008499
  (NUCLEI SciDAC collaboration), NERRSC Grant No.\ 491045-2011, and
  the Field Work Proposal ERKBP57 at Oak Ridge National Laboratory.
  Computer time was provided by the Innovative and Novel Computational
  Impact on Theory and Experiment (INCITE) program.  TRIUMF receives
  funding via a contribution through the National Research Council
  Canada. This research used resources of the Oak Ridge Leadership
  Computing Facility located in the Oak Ridge National Laboratory,
  which is supported by the Office of Science of the Department of
  Energy under Contract No.  DE-AC05-00OR22725, and used computational
  resources of the National Center for Computational Sciences, the
  National Institute for Computational Sciences, and the Notur project
  in Norway.
\end{acknowledgments}

{\it Note added.} -- Very recently Bogner \textit{et al}
\cite{bogner2014} applied the in-medium
similarity-renormalization-group method to construct non-perturbative
shell-model interactions and applied it to neutron-rich oxygen
isotopes.

\bibliographystyle{apsrev}

\end{document}